# On the Complexity of Attacking Elliptic Curve Based Authentication Chips

Ievgen Kabin, Zoya Dyka, Dan Klann, Jan Schaeffner and Peter Langendoerfer
*IHP – Leibniz-Institut für innovative Mikroelektronik*
*Im Technologiepark 25, Frankfurt (Oder), Germany*

*Abstract*— In this paper we discuss the difficulties of mounting successful attack against crypto implementations when essential information is missing. We start with a detailed description of our attack against our own design, to highlight which information is needed to increase the success of an attack, i.e. we use it as a blueprint to the following attack against commercially available crypto chips. We would like to stress that our attack against our own design is very similar to what happens during certification e.g. according to Common Criteria Standard as in those cases the manufacturer need to provide detailed information. When attacking the commercial designs without signing NDAs, we needed to intensively search the Internet for information about the designs. We cannot to reveal the private keys used by the attacked commercial authentication chips 100% correctly. Moreover, the missing knowledge of the used keys does not allow us to evaluate the success of our attack. We were able to reveal information on the processing sequence during the authentication process even as detailed as identifying the clock cycles in which the individual key bits are processed. To summarize the effort of such an attack is significantly higher than the one of attacking a well-known implementation.

*Keywords*— ECC; ECDH; Secure Authentication; FPGA implementation; side channel analysis (SCA) attacks; electromagnetic analysis; horizontal attacks, NXP A1006, Infineon Trust B.

## I. INTRODUCTION

In recent years the number of networked devices and the need for machine to machine communication were increasing dramatically and the prognoses are that this trend will be aggravated by 5G. In order to ensure correct system behavior security features such as confidentiality, data integrity and authentication are essential. This holds not only true for networked devices but also for complex systems such as telemedicine appliances that are compiled of costly parts form highly renown manufacturers. Asymmetric cryptographic approaches such as RSA and elliptic curve cryptography (ECC) are key when it comes to ensuring data integrity, authentication and non-repudiation. The Diffie-Hellman protocol (DH) was proposed in 1976, it exploits modular exponentiation in finite fields to share a secret between two participants. DH allows mutual authentication of both participants or of only one of the participants. This is essential to ensure data integrity and non-repudiation. The modular exponentiation is a time consuming and processing intense operation. In [1] Neal Koblitz proposed to use the elliptic curve (EC) point multiplication instead of the modular exponentiation. The EC-based approach for generation of a shared secret is called Elliptic Curve Diffie-Hellman (ECDH).

For the currently standardized and used cryptographic algorithms it is common sense that they cannot be broken by cryptanalysis, but the situation changes fully when side channel attacks (SCA) are taken into account. With the advent of the Internet of Things a lot of devices are no longer physically protected so that side channel attacks need to be taken into account. The main operation in ECC is the multiplication of an EC point *P* with a scalar k, denoted as *kP* operation. The scalar *k* is the private key if EC-based authentication is performed. The security of cryptographic approaches is based on the secrecy of the private key. Thus, the goal of SCA attacks is to reveal the private key *k* using any information available about the process of the *kP* calculation. For example, the time of *kP* executions, the energy consumption or the electromagnetic emanation measured while performing the private key operation can be analysed to reveal the key.

The EC point multiplication can be implemented in hardware. There a highly specialized implementations available that support only authentication e.g. in form of authentication chips. The EC over binary extended fields $GF(2^n)$ such the standardized NIST curves [2] B-163, B-233, B-283 are especially suitable for hardware implementations. This is due to the fact that the field operations such as addition and multiplication do not require carry-bit propagation. This reduces the execution time of cryptographic operations as well as the area of authentication chips significantly leading to low manufacturing cost for authentication chips. Authentication chips from different manufacturer are available on the market, for example [3]-[4]. Their costs are up to 60 cents. The typical application is the authentication of devices or their parts, for example confirmation of the originality of printer cartridges, electronic accessories such as AC/DC adapters, cables, keyboards, docking stations, batteries, digital headsets, electronic cigarettes etc.

Resistance against SCA attacks is a very important feature for authentication chips. If the scalar *k* used in the *kP* operation can be extracted by an attacker, the attacker can control the authentication of the attacked device and falsify its identity.

In this paper we report about the challenges of attacking implementations of cryptographic operations for which only limited information is available. We decided to attack authentication products from NXP and Infineon for the following reasons. On the one hand these devices are providing core functionality to prevent product piracy and by that improve product safety, i.e. they serve an important issue. On the other hand both implementations are using elliptic curve cryptography for implementing their authentication means. As we have our own implementation of the *kP* operation we are at least familiar with the core functionality.

For both commercial designs investigated here, only limited information is publicly available. Implementation details such as the number of clock cycles needed to complete a *kP* operation, time taken for data exchange etc. are available when attacking your own design. If a manufacturer wants to get a certification for its security implementation the manufacturer has to provide not only the above mentioned details but many other implementation details to the certifier. For an attacker the situation is fully different, no manufacturer will provide such detailed information. Even worse from the attackers point of view is that for authentication products attackers cannot feed the designs with a known scalar to verify the number of successfully revealed bits as a kind of "calibration" of the attack against the secret stored in the chips. So, the granularity of the success definition is binary: all i.e. 100 per cent or nothing i.e. 0 per cent, even though eventually just a single bit of the secret not revealed correctly. The issue that an attacker cannot use known scalars to learn more about the implementation is due to the following two facts:

1. Both authentication chips support only authentication, i.e. in contrast to Elliptic Curve Digital Signature Algorithm (ECDSA) there are no *kP* operations executed on externally provided scalars.
2. The implementation of the *kP* operation used by the software part of the authentication protocol differs significantly from the hardware implementation so that it cannot be used to learn anything about the hardware implementation.

We attacked commercial authentication chips in order to verify if an attacker can successfully extract keys from these designs using public available information. We consider this especially interesting as highly innovative attacks are published every now and then, so that designs implemented before the attack was known may be vulnerable. Here we focus on horizontal address bit DPA as the first one was published when the attacked designs were already available on the market. Please note that our knowledge about the attacked designs is limited to public available information, while an attacker most probably will not hesitate to sign the required NDAs to get more information, and then to also exploit this additional information in the attack(s) launched. Thus, the fact that we were not able to extract a fully correct key does not mean a highly motivated attacker will also not be successful. For our research describe here we used commercial products that are not certified. But, we assume that they despite that, are implemented carefully and that appropriate countermeasures are used to harden the devices.
Our key findings are, even with the limited information available we could successfully understand major parts of he implementations i.e.:

- Isolation of the *kP* operation i.e. we managed to identify all initialization und communication steps
- Time and number of clock cycles needed to complete a *kP* operation.

This paper is an extended version of the paper entitled "On the complexity of attacking Commercial Authentication Products" presented in 2019 on the NTMS Workshop on "CyberSECurity on HARDware" [5]. This version extends the conference paper by giving the following additional information:

1. We provide a very detailed description of both our own kP implementation, so that in principle a kind of whitebox cryptography becomes feasible. Second we provide insight into our own attack and use different target platforms i.e. simulation, ASIC and FPGA to discuss the difficulties when applying our own attack. The key point here is that from simulation to FPGA we have less and less detailed information to mount the attack. But, in any case attacking our own implementation shall provide a blueprint/guideline how to apply the attack on external here commercial designs.
2. We discuss the certification procedures corresponding to The Common Criteria Standard in order to show how much information is needed for certification to highlight the complexity of the our attack compared to the certification procedure. It also kind of provides an impression what a potential use will get with respect to the security level from the certificate received.
3. We provide more detailed information on the chips attacked, in the sense of additional material such as x-ray pictures, a proper discussion of how and where we retrieved information to run the attacks was retrieved from, ranging from publicly available sources to own experiments. We also included first attempts of attacking a new product i.e. NXP A1007 for which even less information than for the A1006 is available.

The rest of this paper is structured as follows. The authentication based on Elliptic Curve Diffie-Hellman (ECDH) approach as well as attack performed against our own design including its implementation details are described in section II. Our analysis of the commercial products is given in section III. Attack results are summarized in section IV. The paper closes with conclusions.

II. ATTACK DESCRIPTION USING OUR OWN DESIGN: WHITE BOX CRYPTOGRAPHY

*A. Attacked algorithm*

Based on the ECDH approach it is possible to authenticate a participant using the public key of the participant. For example, Alice can authenticate Bob using the knowledge of the Bob's public key. To do so Alice sends a request for his certificate to Bob. Bob answers to Alice, i.e. he sends to Alice his certificate that contains the public key of Bob signed by trusted authority. Alice verifies the certificate. Thereafter Alice is sure that Bob is the owner of the public key ***Pub**$_B$*. To be sure that her communication partner is really Bob, Alice generates a random number ***r***, calculates two elliptic curve point multiplications: ***Q**=r·**Pub**$_B$* and ***R**=r·**G***, where ***G*** is the base point of the EC and sends point ***R*** to Bob (see TABLE I. ).

TABLE I. AUTHENTICATION BASED ON ECDH APPROACH.

| *A* (Alice) | *B* (Bob) |
|---|---|
| knows point *G* of an EC | knows point *G* of an EC |
| knows public key of Bob $Pub_B$ | B is owner of: $k_B$; $Pub_B$ |
| • generates a random *r* | |
| • calculates $Q = r \cdot Pub_B$ | |
| • sends to *B*: $R = r \cdot G$ | • receives *R* |
| • receives $Q_B$ | • sends to *A*: $Q_B = k_B \cdot R$ |
| • if $Q = Q_B$ authentication is **ok** | |

Bob calculates the elliptic curve point multiplication $k_B \cdot R$, where *R* is the EC point received and $k_B$ is his private key and sends the result of the calculation $Q_B$ to Alice. Alice compares the received EC point $Q_B$ with the calculated point *Q*. The authentication passes if both points are equal. Please note that $Q_B = k_B \cdot R = k_B \cdot r \cdot G$ and $Q = r \cdot Pub_B = r \cdot k_B \cdot G$, i.e. the EC point *Q* is equal to the EC point $Q_B$ only if an owner of the key pair $k_B$; $Pub_B$ performs all the calculations. It can be Bob or an attacker who successfully revealed the Bob's private key, whereby Bob can be fully sure that only he is in possession of the private key $k_B$. This is due to the fact that SCA attacks can be performed unnoticed by the attacked device/person. Assuming an attacker has physical access to Bob's device. The attacker can send one or many authentication requests instead of Alice and measure for example the electromagnetic emanation of Bob's device during the $k_B \cdot R$ operation. The attacker can collect the measured electromagnetic traces for visual inspection or statistical analysis with the goal to reveal the key $k_B$. If an attacker revealed the private key of Bob, the identity of Bob is stolen. Thus, the resistance of ECDH-based authentication protocols against SCA attacks is based on the resistance of the implementation of the *kP* algorithm against these attacks.

The algorithm that is world-wide most often used for implementing the EC point multiplication is the Montgomery *kP* algorithm using Lopez-Dahab projective coordinates, if the EC is defined over an extended binary field $GF(2^n)$ [6]. This algorithm allows to do the calculation with just the *x*-coordinate of the input EC point. This helps to reduce the execution time and energy consumption of the authentication significantly. Recovering and sending of the *y*-coordinate of the results can be saved additionally. Algorithm 1 shows the mostly referenced variant of the *kP* algorithm published in [7]. The *l* bit long scalar *k* is processed from its most significant bit $k_{l-1}$ to its least significant bit $k_0$, i.e. left-to-right.

---

**Algorithm 1**: Montgomery *kP* using projective Lopez-Dahab coordinates

Input: $k = (k_{l-1} ... k_1 k_0)_2$ with $k_{l-1} = 1$, $P=(x,y)$ is a point of EC over $GF(2^n)$
Output: $kP = (x_1, y_1)$
1: $X_1 \leftarrow x$, $Z_1 \leftarrow 1$, $X_2 \leftarrow x^4+b$, $Z_2 \leftarrow x^2$
2: **for** $i=l-2$ **downto** $0$ **do**
3:    **if** $k_i=1$
4:       $T \leftarrow Z_1$, $Z_1 \leftarrow (X_1Z_2+X_2T)^2$, $X_1 \leftarrow xZ_1+X_1X_2TZ_2$
5:       $T \leftarrow X_2$, $X_2 \leftarrow T^4+bZ_2^4$, $Z_2 \leftarrow T^2Z_2^2$
6:    **else**
7:       $T \leftarrow Z_2$, $Z_2 \leftarrow (X_2Z_1+X_1T)^2$, $X_2 \leftarrow xZ_2+X_1X_2TZ_1$
8:       $T \leftarrow X_1$, $X_1 \leftarrow T^4+bZ_1^4$, $Z_1 \leftarrow T^2Z_1^2$
9:    **end if**
10: **end for**
11: $x_1 \leftarrow X_1/Z_1$
12: $y_1 \leftarrow y+(x+x_1)[(X_1+xZ_1)(X_2+xZ_2)+(x^2+y)(Z_1Z_2)] / (xZ_1Z_2)$
13: **return** $(x_1, y_1)$

---

The *kP* calculation corresponding to Algorithm 1 is a sequence of mathematical operations with elements of the extended binary Galois field $GF(2^n)$: field multiplications, squarings and additions. The number of the mathematical operations for processing a key bit '1' is the same as for processing a key bit '0', i.e. it does not depend on the value of the processed key bit. This is the reason, why the Montgomery *kP* algorithm using the Lopez-Dahab projective coordinates for representing EC points is referred in the literature as resistant against simple SCA attacks, i.e. against single-trace attacks using a visual inspection of the measured trace for revealing the key.

Please note that in an SCA-aware implementation not only the number but also the sequence of all operations, including the operations with registers (data storing as well as the reading), has to be the same for each processed key bit value. The main goal of designers is to make the shape of the processing of a key bit value '1' indistinguishable from the shape of processing of a key bit value '0' in the measured trace(s). Algorithm 1 shows a way to calculate *kP*. The number of registers, blocks, clock cycles for each operation, even the number of the field multiplications, as well as the parallelization of the calculations, and many other implementation details are usually defined by designers according to predefined optimization criteria. Applying the regularity, balancing and atomicity principles while designing an implementation helps to increase the resistance of the Montgomery *kP* algorithm against simple SCA attacks. We implemented our own *kP* design for the NIST EC B-233 applying these principles.

*B. Implementation details of our kP design:*

In this subsection we describe implementation details important when performing attacks.

Our design consists of the following functional blocks:
- a field multiplier;
- a unit for addition or squaring of the field elements, depending on the control signal;
- registers for storing of the inputs, outputs, as well as intermediate results;
- a control unit that manages the sequence of the operations;
- a system muxer – the bus – realizing the data flow between the functional blocks.

A field multiplication of 233 bit long operands takes 11 clock cycles in our implementation: 2 clock cycles for receiving the new multiplicands and 9 clock cycles for the calculation and accumulation of the 9 partial products (including the field reduction) corresponding to the 4-segment Karatsuba multiplication method (MM) [8]. Due to the fact that in our implementation the new multiplicands are always given in parallel to the last two partial product calculations, the field multiplication block is always active in our *kP* design, i.e. it calculates a partial product in each clock cycle of the processing of the key bits in the main loop of the algorithm. The 4-segment Karatsuba MM saves about 44% of the calculation time for a field multiplication compared to the classical multiplication method. This is due to the fact that the latter requires the calculation of 16 partial products[1]. We implemented the calculation of partial products of 59 bit long operands as an operation that is completed in a single clock cycle using the classical multiplication method. In [9] we showed that a field multiplier that is always active is a kind of a noise source, especially if its partial multiplier is implemented using the classical MM, i.e. such a multiplier can increase the resistance of the whole *kP* design against SCA attacks.

Our design implements the Montgomery *kP* algorithm according to [10]. The processing of each key bit in the main loop of the Montgomery *kP* algorithm is implemented with 6 field multiplications, 5 squarings, 3 additions and 11 register operations and takes 54 clock cycles only. We denote the part of a measured trace that corresponds to the processing of a single key bit as a slot. The hardware implementation of the Montgomery *kP* algorithm processes the scalar *k* bitwise, whereby the use of registers in the *kP* algorithm depends on the processed bit value $k_i$ of the scalar *k*. This dependency can lead to the fact that vertical and horizontal differential address bit SCA attacks can be successful. A vertical address bit DPA was reported in 2002 in [11]. A well-known fact is that traditional countermeasures against vertical attacks such as scalar randomization as well as the randomization of the EC point coordinates are not effective against vertical address bit DPA attacks [12].

We revealed the scalar *k* successfully performing a horizontal, i.e. single-trace, attack exploiting the key dependable addressing of registers using the noise-free simulated traces of our old design [13] in 2015. We improved the resistance of our implementation significantly using the horizontal differential analysis attacks as a tool for localizing SCA leakage sources [14]. The statistical analysis can be done using the Pearson correlation coefficients, or other statistical approaches, for example a comparison to the mean approach [15]. In the next subsection we describe shortly this attack against our *kP* design with the goal to show how easy and effective horizontal differential attacks against Montgomery *kP* algorithm implementations can be.

### C. Attack description

We synthesized the *kP* design using IHP's 250 mn gate library [16] for a clock cycle of 50 ns i.e. for a maximum working frequency of 20 MHz. We used the "*compile*" option in the Synopsys Design Compiler (version K-2015.06-SP2) to perform default area optimization. The processing of an *l*=233 bit long scalar *k* requires about 1300 clock cycles, whereby only *l*-2 bits of the scalar are processed in the main loop of the algorithm. We simulated the power consumption of the *kP* designs after layout using the Synopsys PrimeTime suite [17]. The *kP* operation was executed using a randomly generated 232 bit long scalar *k* and a randomly selected point *P*. The simulated traces are noiseless. We compressed the simulated power trace with the goal to simplify the analysis, i.e. we represented each clock cycle using only one value – the average power value of the clock cycle. Fig. 1 shows the same parts of the compressed simulated traces of the *kP* execution simulated after synthesis (orange line) and after layout (blue dashed line).

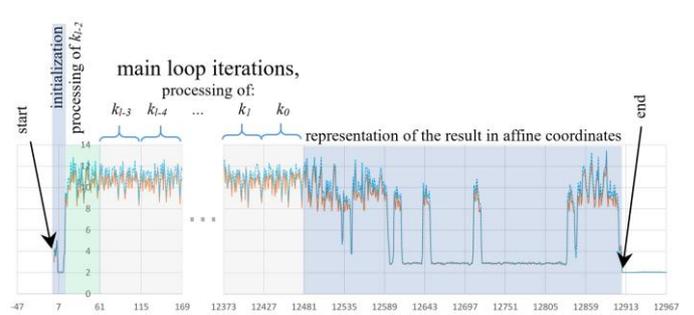

Fig. 1. A partss of compressed traces of a *kP* execution simulated after synthesis (orange line) and after layout (blue dashed line) for our *kP* implementation.

Fig. 1 shows the start of the *kP* operation, the initialization phase of the Montgomery *kP* algorithm (see line 1 in Algorithm 1), the shape of the processing of the second most significant bit $k_{l-2}$ before the main loop (corresponding to Algorithm 2 in [10]), the shape of slots in the main loop as well as the shape of the calculation of the affine coordinates of the result. In our analysis we concentrated only on revealing the key bits processed in the main loop of the algorithm. For the 232 bit long scalar *k* there are (232-2)·54=12420 clock cycles. The main idea of the analysis is to distinguish parts of the trace corresponding to the processing of key bits '1' from parts corresponding to the processing of key bits '0'.

Designers can use their knowledge about the processed scalar *k* to split the analysed trace into two sets of slots, one representing the processing of '1' and the second one representing the processing of '0'. The distinguishability of the '0'-slots from the '1'-slots is than the measure to evaluate the resistance of the implementation against selected horizontal attacks such simple SCA or differential address bit DPA. In an ideal case from the designer's point of view these two sets are not distinguishable, i.e. the mean shape of all '0'-slots and mean shape of all '1'-slots are for example equal or the shapes of all slots are fully random. Designers can apply for example Welch's test [18] to a big set of *kP* traces measured with different but known scalars *k* with the goal to evaluate statistically the distinguishability of the '0'-slots and the '1'-slots.

Assuming that '0'-slots differ (even slightly) from '1'-slots, an attacker can exploit the null hypothesis to reveal the key. He can calculate the mean shape "***mean***" of all slots, for the whole trace, without any knowledge about the key bit value processed in each of the slots. The attacker can try to distinguish the '0'-slots from the '1'-slots based on the assumption that the ***mean*** slot is between both sets i.e. it separates both sets into the set of '0'-slots and the set of '1'-slots. Thus, the attacker can determine key candidates by sample-wise comparing the ***mean*** slot with each $i^{th}$ slot of the measured trace. He compares the sample value with number *j* in the $i^{th}$ slot to the sample value with number *j* in the ***mean*** slot. If it is smaller than in the ***mean*** slot, the the $i^{th}$ bit of the $j^{th}$ key candidate is assumed equal to '1'else it is assumed equal to '0'. By applying this *comparison to the **mean*** to all samples of all slots the attacker extracts *j* key candidates. Calculating the EC point multiplication *kG* (where *G* is the base point of the EC) and comparing the results with the public key of the attacked person/device the attacker can conclude if one of the extracted key candidates is equal to the processed scalar $k$ or not.

---

[1] $44\% = \dfrac{16-9}{16} \cdot 100\%$

Evaluating the success of an attack is by far simpler for the designers, as they know the processed scalar $k$ and can compare each of the key candidates bitwise with the scalar $k$ and use this to calculate the success rate and to determine exactly which bits of the scalar were determined correctly. We express the success rate of the attack as relative correctness of each key candidate, denoted further as $\delta$. The relative correctness of a key candidate is the relation of the number of correctly revealed bits to the number of all key bits processed in the main loop of the attacked algorithm.

Important for the attacks using the *comparison to the* **mean** approach is the knowledge about implementation details such as the start, the end and the duration of each slot. An additional assumption is that all the slots have the same duration. In comparison to designers attackers do not have any knowledge about the start and duration of slots i.e. they even need to speculate about these simple but very important details. A wrong separation of the measured $kP$ trace into slots can reduce the attack success significantly.

### D. Attacking kP in IHP 250 nm technology

We analysed the simulated power traces of our $kP$ design before manufacturing the ASIC as described above. We compare each key candidate with the scalar $k$ bitwise and express the number of correctly extracted bits $\delta$ in per cent. The success of our attack represented as the correctness of each of the key candidates is shown in Fig. 2. The result of the analysis attacking the compressed power traces simulated after synthesis (i.e. before layout) is given by the orange line. The dashed blue line shows the attack success against the trace simulated for the design after layout.

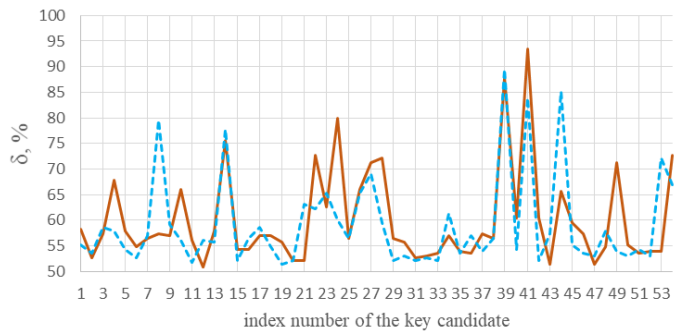

Fig. 2. Results of the *comparison to the mean* horizontal attack against power traces of the *kP* execution simulated after synthesis (orange line) and after layout (dashed blue line).

Due to the fact that the synthesis does not take into account a physical placement of the cells, the signal delays and many other parameters, the shape of the power trace simulated after the synthesis (i.e. before the layout) differs from the shape of the power trace simulated after the layout. This explains the differences in the attack success (see Fig. 2).

After layout and post-simulation of the *kP* power traces our design was manufactured as a 2.5 mm times 1.1 mm silicon chip die using the IHP 250 nm cell library SGB25V [16]. The maximum operating frequency achieved was as high as 20 MHz. Finally, the ASIC was bonded to a printed circuit board (PCB) as shown in Fig. 3.

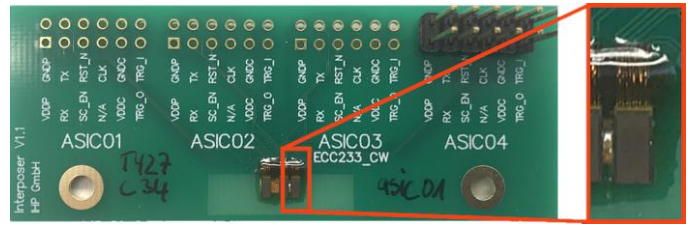

Fig. 3. IHP *kP* accelerators produced in IHP's 250 nm technology bonded to a PCB and a zoomed in die.

We measured an EM trace of the *kP* execution using the measurement setup shown on Fig. 4. It consists of a LeCroy WavePro 254HD oscilloscope and a near-field micro probe MFA-R 0.2-75 from Langer [19] placed into the Langer 4-Axis (3 axis + rotation) Positioning System ICS 105 [20].

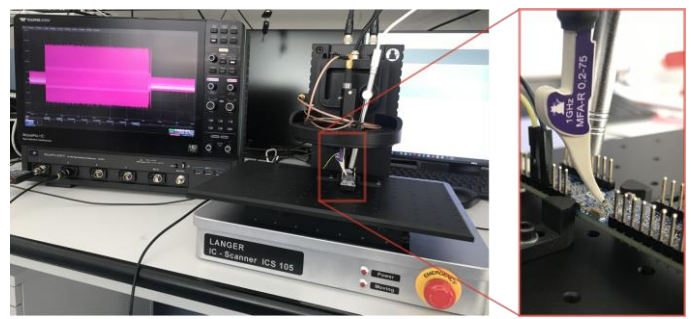

Fig. 4. Our measurement setup.

The measured EM trace of the *kP* execution is shown in Fig. 5.

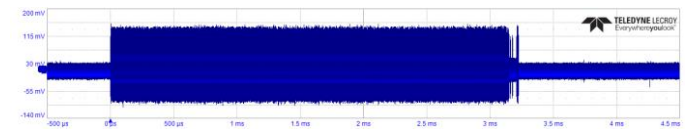

Fig. 5. The measured EM trace of the *kP* execution.

Due to the fact that the simulation data obtained for the design before manufacturing the ASIC contains information about the activity of each single block of the design as well as the value of intermediate variables, which can be analysed individually, the design can be understood in detail very good. This knowledge and understanding can be applied for the analysis of traces measured on the manufactured ASIC. Using this knowledge we determined the start of the 1$^{st}$ slot in the measured trace that we defined also as the start point for the compression of the trace for successive analysis. Without this knowledge the task to separate the measured trace into slots is very complex. We compressed the measured trace calculating the compressed value of a clock cycle as the sum of squares of values measured within the clock cycle. We analysed the compressed trace using the comparison to the mean in the same way as for the compressed simulated traces. Each slot in the compressed trace consists of 54 samples. Using profiles of all slots in the compressed trace we calculated the *mean* slot of the trace. It also consists of 54 values. For each $1 \leq j \leq 54$ we obtained one key candidate corresponding to the attack description in section II-C. Thus, we obtained 54 key candidates attacking the electromagnetic trace of our *kP* design. We compared each key candidate obtained with the processed scalar *k*. We calculated for each key candidate its relative correctness δ as the ratio of the number of correctly revealed bits in the $j^{th}$ key candidate to 230 which is the number of key bits processed in the main loop of the Montgomery *kP* algorithm. Fig. 6 shows the success of the attack against our *kP* design using the measured EM trace (see violet line) in comparison to the attack using the *kP* trace simulated after layout (see blue dashed line).

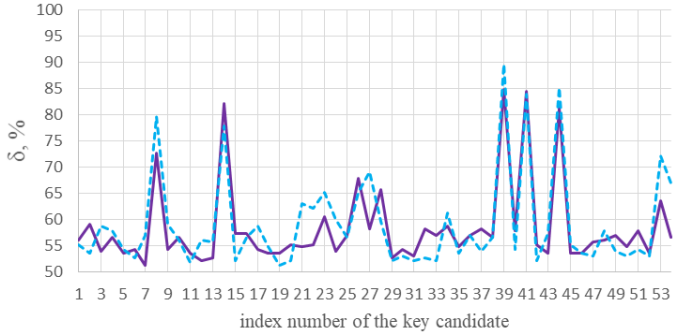

Fig. 6. The results similitude of the *comparison to the mean* attack performed against measured electromagnetic trace (violet line) of the *kP* execution and a simulated power trace obtained after layout (blue dashed line).

The result of the analysis attacking the power trace simulated after layout and attacking the measured EMT of the manufactured chip are very similar. Thus, the evaluation of the design resistance in the earlier designing phase, i.e. before the chip manufacturing, is possible and reasonable.

*E. Attacking our kP design running on FPGA*

We ported our design to a Xilinx Spartan-7 FPGA (Cmod-S7 board from Digilent [21]) running at 100 MHz. The attacked FPGA is highlighted by a red rectangle on the board shown in Fig. 7-a).

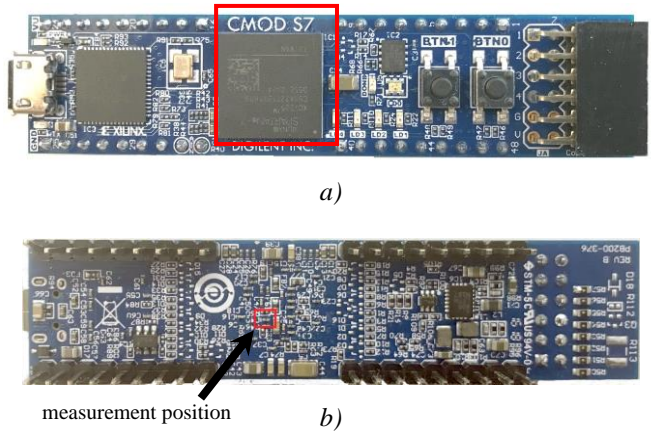

Fig. 7. Digilent Cmod S7: Breadboardable Spartan-7 FPGA Module: *a)* – front side with the attacked FPGA; *b)* – back-side with the measurement position.

The execution time for a *kP* operation is about 130 μs, i.e. about 7700 *kP* operations per second. All measurements performed in this paper were obtained using the same measurement setup. We captured electromagnetic traces during the authentication process using the measurement setup shown in Fig. 7. The near-field microprobe was positioned at the power decoupling capacitor C40 [22] (see Fig. 7-*b)*) close to the attacked chip. The measured trace and parts zoomed in are shown in Fig. 8. The part zoomed in consists of 50 clock cycles i.e. it is about one slot long.

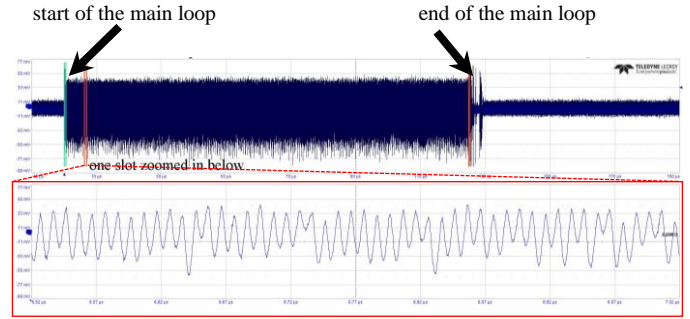

Fig. 8. . Electromagnetic trace measured during the *kP operation*.

As it can be seen the shape of the whole EM trace measured on the Spartan 7 FPGA is quite similar to the one obtained for the ASIC (see Fig. 5). Similar to the EM trace measured on the ASIC the trace measured on the FPGA cannot be easy separated into slots, i.e. the start point of the processing of key bits in the main loop of the implemented algorithm as well as the duration of the processing of each key bit are by far not obvious. Using our knowledge about the design and the simulation trace as auxiliary material we were able count how many clock cycles the design needs before the $1^{st}$ main loop iteration starts. Thus, we counted this fixed number of clock cycles from the beginning of the *kP* operation to determine the start point for the analysis. Beginning from the determined start point we compressed the measured trace at the same way as for the ASIC. We analysed 230 compressed slots using the comparison to the mean in the same way as for the ASIC. Fig. 9 shows the success of the attack.

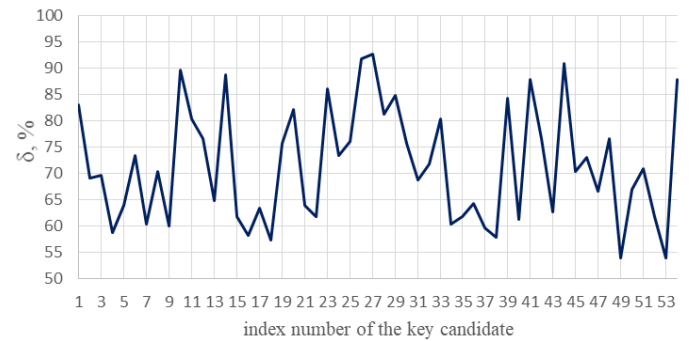

Fig. 9. Results of the *comparison to the mean* horizontal attack analysing the EMT trace measured during a *kP* execution on an Spartan-7 FPGA.

The highest key correctness δ obtained is 92.6% i.e. only 17 of the 230 key bits were revealed incorrect. If their positions are known – the complete key can be brute forced by performing $2^{17}$= 131072 point multiplications.

Please note that evaluating the attack success is easy only for designers, or if a *kP* execution as part of an ECDSA verification [23] is attacked. For the latter a scalar can be provided by the attacker with which then the *kP* execution is performed. As attackers, i.e. without the knowledge of the processed scalar *k*, we can evaluate the extracted scalars *k* only using the calculation *kG*. Thus the success of the attack can be evaluated either with '*yes*' if the result of the *kG* calculation is equal to the public key of the attacked person/device, or with '*no*'. Due to the fact, that no key candidate is equal to the real scalar *k* in our attack, the attack was not successful from the attacker's point of view.

Table II summarizes the results of our attack against our own design: simulation for the IHP 250 nm technology, manufactured ASIC in the IHP 250 nm technology and an FPGA implementation.

TABLE II. AUTHENTICATION BASED ON THE ECDH APPROACH.

| *kP* implementation attacked | | analysed trace | F, MHz | attack success as highest key correctness (corresponding clock cycle) | Number of key candidates with a correctness of more than 80% |
|---|---|---|---|---|---|
| simulation | synthesis | PT | 20 | 93.5% (41) | 3 |
| | layout | PT | 20 | 89.6% (39) | 3 |
| ASIC | | EMT | 4 | 84.8% (39) | 4 |
| FPGA | | EMT | 100 | 92.6% (27) | 15 |

The data in Table II demonstrates clearly that the resistance of a cryptographic implementation against SCA attacks depends on the target platform and/or targeted operating frequency. Despite the fact, that the same VHDL code was running and the same input data were processed in all experiments, the success of the attacks is different. The main difference is the number of the key candidates revealed with a high correctness. For the FPGA implementation running at 100 MHz 15 of 54 key candidates were extracted with the correctness of more than 80% i.e. at least 184 of 230 bits of the processed scalar *k* were revealed correctly. The best key candidate was revealed attacking the FPGA implementation with a correctness of 92.6% i.e. only 17 key bits were revealed incorrect. The knowledge about which key bits were incorrectly revealed can be exploited to speed-up brute forcing a fully correct key. Please note that no key candidate was revealed with a correctness of 100% in any of our experiments discussed here and the positions of the wrongly revealed key bits are unknown. This means – from the attacker point of view – that the attacks were not successful. But from designer's point of view revealing a key with a correctness of about 93% means that the design has a strong SCA leakage source and is potentially vulnerable to the performed SCA attack.

As mentioned above, our *kP* design is a strongly balanced implementation of the Montgomery *kP* algorithm. Additionally, the always active field multiplier is a kind of SCA protection due to the high fluctuation of its power consumption [9]. Traditional countermeasures against vertical data bit differential attacks [24] were not implemented in our design due to the fact, that the success of the horizontal address bit differential attacks does not depend on the:

- EC point *P* processed in the analysed *kP* calculation [12] i.e. EC point blinding and randomization are not effective against horizontal (single trace) attacks
- scalar *k* processed in the *kP* calculation [12]. If the scalar, randomized corresponding to the key randomization approaches [24], will be successfully revealed, the attacker can use the revealed randomized scalar instead of the original scalar *k*.

The success of the horizontal attacks depends directly on the attacker's knowledge of the following details:

- algorithm implemented
- start and end of the *kP* operation;
- start, end and duration of slots;
- stable clock frequency;
- processed scalar to evaluate the attack success and to learn about wrongly revealed bit positions;
- countermeasures implemented.

Hiding this information, stabilization or randomization of the power consumption of the design, randomized frequency of the clock signal – all these measures can reduce the success of the horizontal attacks significantly. Thus, SCA attacks against each new external design start with collecting information about implementation details of the device to be attacked.

### III. INDUSTRIAL AUTHENTICATIONS PRODUCTS

There is no widely accepted metrics for the security level a certain implementation provides. At best the key length is considered a reasonable indicator, in the sense that a longer key provides a higher security. But the issues with this metrics is that Side Channel Attacks are not really taken into account. We agree that increasing the number of key bits to be revealed in some way increases the effort for a potential attacker, but whether or not SCA will be successful or not solely depends on the quality of the implementation i.e. it can be fully decoupled from the key length. So, in case of a sloppy implementation the extra effort may be fully negligible.

The Common Criteria Standard (CCS) [25] is one of the most often applied means to evaluate the security level of a certain product. It proposes a kind of a decision metric that is applicable to many specific cases. In order to ensure proper application of the CCS a guideline is provided which needs regular updates. For example the ECC evaluation guide [26] describes attacks against which an ECC chip shall be resistant on "only" 30 pages, but does not exclude to extend the list at any time.

The CCS evaluation flow consists of the following 3 steps: understanding the security design, vulnerability analysis and testing. The pass/fail verdict is based on the estimated attack potential, whereby the public accessible information about the implementation details of the attacked design can benefit the attack success and – consequently – is a negative factor when estimating the design resistance level. Time spent for the attack, expertise level of the attacker/evaluator as well as the equipment needed are additional factors for the evaluation of the resistance of cryptographic implementations.

When certifying a security implementation corresponding to CCS, developers have to provide all design information for the evaluation, so all implementation details are known to the evaluator meaning the evaluation is a white box cryptography analysis. Even if the knowledge about implementation details, including countermeasures implemented against SCA attacks, is available the certification takes quite long about 1 year and is costly about 200,000-350,000 Euros [27].

Many security chips currently available on the market are not certified according to CCS, for example ECC-based authentication chips from Infineon SLE95250 [4] or NXP A1006 [3]. Please note that the fact of being not certified provides almost no hint concerning the security level of a certain chip. The manufacturer may have decided not to go for certification for timing and/or financial issues while experienced designers do their very best to avoid all known pitfalls and in-house test teams run all essentially needed tests. The issue here is that missing certifications and/or incomplete information on the certification result for certified chips, kind of shift the problem to teams that try to build more complex devices using crypto chips as building parts. What we mean here is that these teams cannot take informed design decisions as essential information is missing.

The Montgomery *kP* algorithm is to some extend the de-facto standard algorithm for most commercial authentication products. This mean known countermeasures against the vertical and horizontal address bit DPA such as the randomized addressing of registers [28] or the randomization of the main loop of the Montgomery *kP* algorithm [29]-[30] or developing a regular schedule in which the blocks are addressed [14] have to be implemented. The resistance of *kP* designs can additionally be increased by depending on the activity of the field multiplier. To exploit the field multiplier as a countermeasure different multiplication methods have to be combined [31]. We did not find any information that such a countermeasure was applied in any of the researched commercial chips. We found only classical countermeasures such as the randomization of coordinates of EC points [32].

Each of the following subsections is structured as follows. First, we present the public available information we gathered as a basis for our attacks. Please note the more information an attacker has at hand the easier the attack can be. Afterwards we present the measured EMTs of the commercial chips we researched. At the end of the subsections we discuss the complexity of applying our software for horizontal differential SCA to the measured traces as well as the results achieved.

### A. Infineon Optiga Trust B

#### 1) Basic Information from its specification

This hardware based security solution is available since 2015 and can be used in a wide range of applications and is designed for easy integration into embedded systems. The main area of its application is a one way authentication of replacement parts with a special focus on batteries to help system and device manufacturers to ensure authenticity, integrity and safety of their original products. It can also be applied in devices for IoT for IP and PCB design protection, and in medical and diagnostic equipment.

According to [4] main features are:

- Single-Wire I/O Interface
- High level of Security – 131 bits
- 163 bits OPTIGAT Trust B Digital Certificate (ODC)
- Single Supply Voltage Support (From 2.0V to 5.5V)
- 96 bits Unique Chip Identification number
- Max Response Computation Time ECCE131 34.0 ms

#### 2) Our measurements

The evaluation board for an OPTIGA Trust B is shown in Fig. 10. The secure authenticator is an IC marked as U308 on the board (see the top left part of the Optiga evaluation board, green circle). The measurement position is shown in Fig. 10 by the blue arrow.

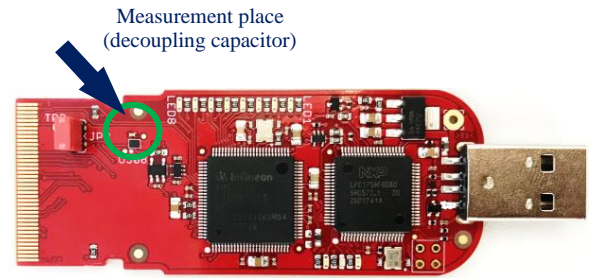

Fig. 10. OPTIGA Trust B SLE 95250 Evaluation Kit board.

The functionality evaluation is performed using a software from Infineon. Its GUI and the 9 steps to be performed for authentications are shown in Fig. 11.

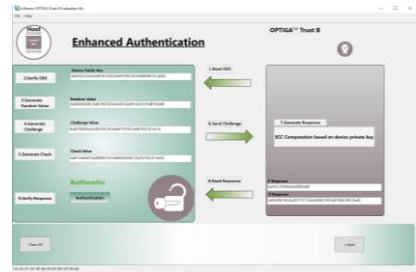

1. Read ODC
2. Verify ODC
3. Generate random value
4. Generate Challenge
5. Generate check
6. Send Challenge
7. Generate Response
8. Read Response
9. Verify Response

Fig. 11. Graphical user interface of the evaluation software of the Infineon OPTIGA Trust B secure authenticator.

The trace shown in Fig. 12-*(a)* corresponds to the execution of all authentication steps. A part of the trace on the left side matches to the reading of the Optiga Digital Certificate (ODS) from the chip (Step 1). This step takes about 500 ms. Three regions of the trace from the right side are shown zoomed in in Fig. 12-*(b)* and correspond to the transmission of the challenge from the host device to the chip (Step 6), response generation (*kP* operation, Step 7), and transmitting the response back to the host device (Step 8). The steps 6-8 require 120 ms according to the measurements. In Fig. 12-*(c)* a Response Generation (25 ms) and a zoomed in part representing several clock cycles of the processing are shown. It can be seen that a clock cycle period is equal to 125 ns. So the *kP* design is running at 8 MHz and requires about 200,000 clock cycles for the *kP* calculation. Infineon did not publish any information about SCA countermeasures implemented.

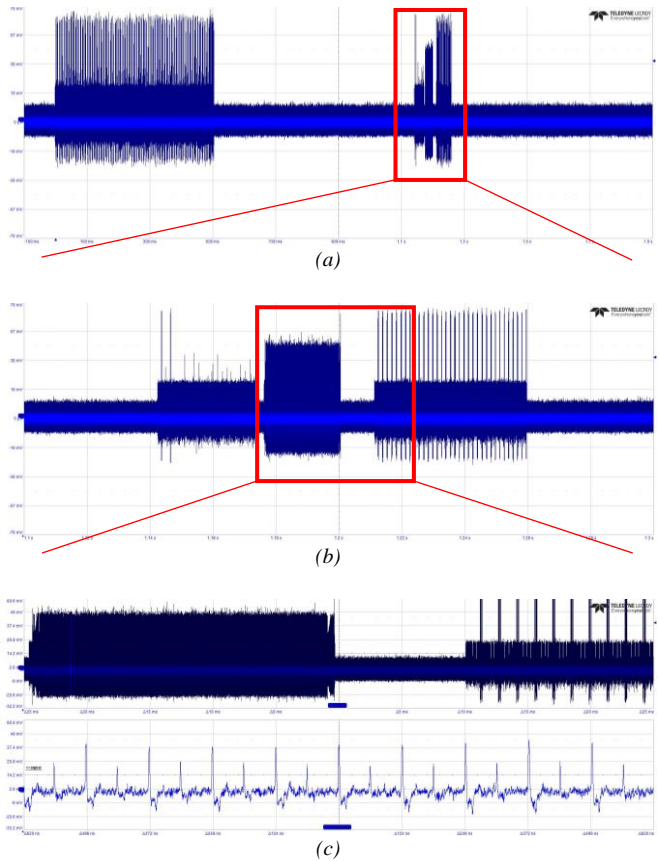

Fig. 12. Electromagnetic trace measured during an authentication executed on the Optiga Trust B authentication chip

### B. NXP A1006 Secure Authenticator

*1) Basic Information from specification*

The A1006 Secure Authenticator [3] for anti-counterfeit applications is a low power and small footprint integrated circuit with protection against various invasive and noninvasive attacks. The chip is available in two different packages – an SOT1189-1/XSON6 plastic package with 6 pads and WLCSP4 4 pads bump Wafer Level Chip flip chip (see Fig. 13). After decapsulation of the chip in plastic case we did a visual comparison with the WLCSP4 chip. We found that layout as well as dimensions (1.03 mm x 0.94 mm x 0.5 mm according to [33]) of the chips look very similar. The description of the A1006 [34], reports that, pin 2 of the plastic chip is not connected (n.c.) but should be connected to ground. We x-rayed the A1006 revealing that this pin is connected see Fig. 14.-a. In this figure all 6 pins of the package are bonded to the die. By performing additional measurements we determined that this pin is not connected internally to the ground. The purpose\functionality of this pin remains unclear for us.

The A1006 datasheets available since 2016 report on the following countermeasures: active and passive shielding memory scrambling and security sensors [34] and the use of randomized projective elliptic curve point coordinates [32]. We were unable to find any information on any additionally applied SCA countermeasures.

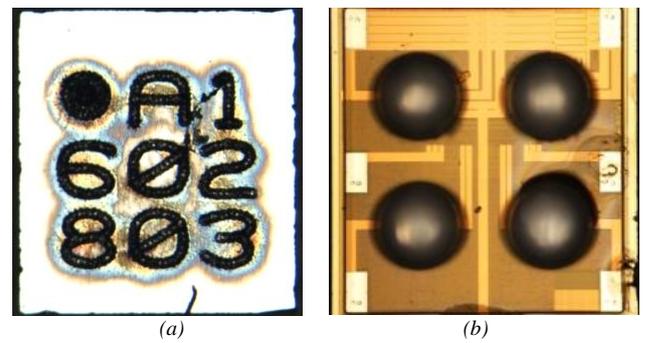

Fig. 13. A top side *(a)* and bottom side with balls *(b)* of an A1006 secure authenticator IC produced as a BGA flip-chip.

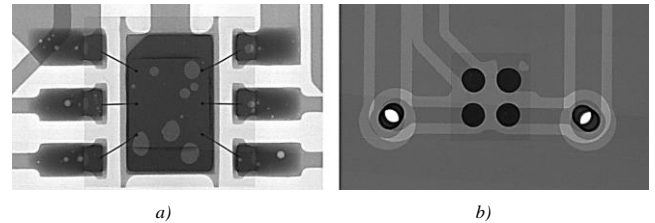

Fig. 14. X-Ray photos of an NXP A1006 integrated circuit in a SOT1189-1/XSON6 plastic package *a)* and a WLCSP4 Wafer Level Chip b).

Key Features according to [3] and [32] are:
- asymmetric authentication protocol based on NIST B-163 elliptic curve (see Fig. 15 )
- Digitally signed certificates using 224-bit ECDSA and SHA-224 digest hash
- 64 bit unique identifier
- 4kbit EEPROM
- security features include TRNG, active shielding, security sensors.
- Power supply range from 1.8 V to 3.3 V
- OWI and I2C Interfaces
- Transmission of the challenge/response is performed using EPIF (error protected isomorphic field) coordinates [35]

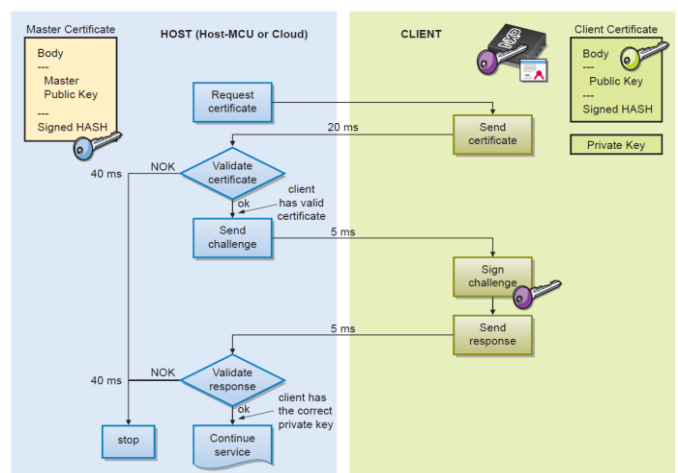

Fig. 15. NXP A1006 authentication flow. The image is taken from [34].

The unique identifier is assigned during wafer production and can be used to trace an A1006 IC all of the way back to its position on the wafer it came from. The 64 bit length provides guaranteed unique values for 10 years of production across all NXP IC's.

4 kbit EEPROM is splitted into 4 regions 1 kbit each according to the Fig. 16.

| 1st Cerificate area | 2nd Cerificate area | User Memory | System Memory |

Fig. 16. NXP A1006 EEPROM details according to [36]: 4 kBit splitted into 4 regions 1 kBit each.

- The first region is read only containing the NXP Certificate.
- The second region contains a user defined certificate that is inserted during the customer manufacturing flow. Once the certificate is written this region is "locked" and read only for the rest of the lifetime of the chip.
- The third region is always accessible (read/write) for user's needs.
- The fourth region is system memory that cannot be accessed by any customer ever.

The NXP certificate contains information regarding the unique identifier of the chip and other customer information as well as a public key. The corresponding private key is stored in the secure storage of the IC and never leaves it. This die individual key pair, certificates and other personal information are inserted into each chip during fabrication process in secure manufacturing facilities of NXP.

*2) Our measurements*

The demonstration kit for an A1006 Secure Authenticator is shown in Fig. 17. It consists of an LPC1115 board, A1006 Sandwich Board and A1006 IC placed into a test socket. The GUI for the evaluation board is shown in Fig. 18.

Full authentication is performed using the following sequence of operations:
1. Hardware initialization
2. Reading of the unique identifier of the A1006
3. Reading of the compressed certificate of the A1006
4. Parsing and decompression of the certificate into X509v3 format
5. Verification of the certificate and its signature
6. Generation of a challenge message
7. Precomputation of the challenge response using the A1006 public key
8. Sending the challenge message to the A1006
9. Response computation performed on the A1006
10. Reading the challenge response from the A1006
11. Verification of the received challenge response

We measured the electromagnetic traces (see Fig. 19) during a full authentication of A1006 chip on a capacitor placed between power supply and ground lines. In Fig. 19-*a* fragments 1 to 4 in our opinion correspond to steps 2, 3, 8, 9 of the full authentication. Fragment 7 corresponds to step 10, the meaning of fragments 5 and 6 is unknown to us.

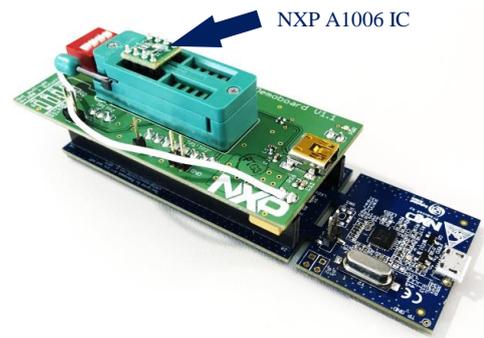

Fig. 17. NXP OM13589JP Demonstration Kit with an A1006 IC (highlighted by an arrow) installed in ZIF DIP test socket.

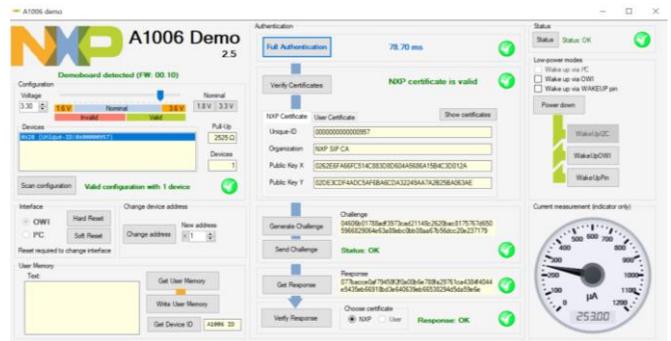

Fig. 18. Graphical user interface for an evaluation of the NXP A1006 secure authenticator. Full Authentication takes about 79 ms.

According to the A1006 Demo Software a full authentication takes up to 80 ms (see Fig. 18). But according to the measured trace it takes about 240 ms. The chip returns a point multiplication result only 130-140 ms after its calculation was finished (see a peak close to 240 ms in the Fig. 19-a). It's not clear whether reading the multiplication result is triggered by the software or whether there is a delay defined by hardware. We plan to investigate this later.

Twenty regular and one short parts of the measured trace shown in Fig. 19-c probably conform to the processing of an up to 163 bits long key. It seems that each of these regular parts corresponds to the bitwise processing of eight key bits.

The clock frequency of the device is low and not stable as can be seen in Fig. 19-c. Only 5% of the clock cycle duration correspond to switching of logic in the IC, the rest is noise. Therefore it is necessary to use an oscilloscope which is able to provide significant oversampling and that can store such a huge amount of measurement samples. All these factors hinder the analysis of measured traces. Adapting software for the horizontal differential analysis of traces is a time-consuming task and was not done. For a motivated attacker it will be not an impediment.

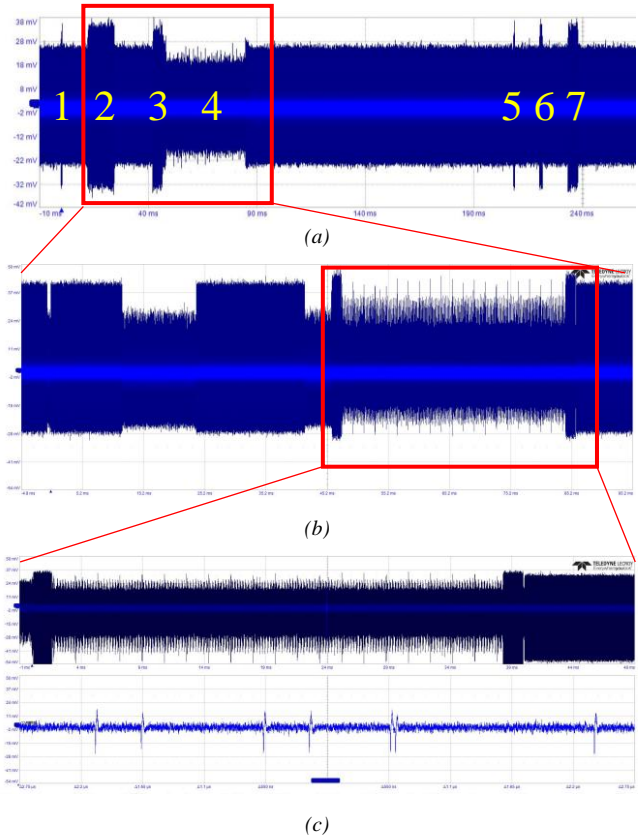

Fig. 19. Electromagnetic traces measured during full authentication *(a)*, part of the first half of the authentication zoomed in *(b)* and generation of an A1006 response *(c)*.

Our attack failed due to the lack of the information about the NXP A1006 and which motivated us to perform additionally some experiments with the NXP A1007 Secure Authenticator. It is a relatively new chip on the market, that was announced in 2018 [37]. Like its predecessor, A1007 is mainly focused on anti-counterfeit protection and proof of origin. It is similar to the A1006, but there is not much information given about it in freely available sources. NXP claims countermeasures against timing attacks as well as simple power analysis, differential power analysis, electromagnetic analysis and protection against differential fault analysis attacks for this chip. For these countermeasures no detailed description is available. The supplied OM67201UL kit that we bought for experiments consists of the NXP LPC11U37H Xpresso v2 board combined with A100x_SHIELD_A3 and an A1006 socket board, where the A1006 secure element was replaced by the A1007. The kit does not include any software for the evaluation in contrast to OM6700 and OM13589 (A1006 Development board and Evaluation board respectively). According to the "Quick Start Guide", it is necessary to follow the link www.nxp.com/A1007-kit in order "to access the full Getting started Guide and download the latest SDK". Unfortunately, the provided link is broken and NXP Technical Support team informed us that "All the information about the A1007 is under NDA and you need to enter Docstore to get/request it".

We tried to use the NXP A1006 demo software with an A1007 inserted into OM13589JP Demonstration Kit, but the device was not detected by the software. Only the C project provided together with the OM6700 Development board for A1006 partially spilt the beans on the device parameters. We were able to execute steps 1 to 4 from the full authentication process described above for the A1006 chip. The certificate obtained from the chip during step 4 did not pass the verification, i.e. steps starting from 5 failed. Nevertheless, from this certificate we know that this chip as well as its predecessor has:

- a 64 bit unique identifier;
- uses NIST B-163 EC-based authentication protocol;
- digitally signed certificates using 224-bit ECDSA and SHA-224 digest hash.

This information we found also in [38].

## IV. ATTACKS SUMMARY

TABLE III. provides an overview of the main information we have revealed about the attacked designs. Please note that all entries in the table are result of our own experiments except of those for which we provide references. The information given on the implementation of the NXP A1006 i.e. that it uses the Montgomery algorithm was deduced from information in [32] which reports on the use of Lopez-Dahab (LD) coordinates which to the best of our knowledge can be used only in combination with the Montgomery algorithm. All other information such as clock frequency, time and number of clock cycles for executing a *kP* operation have been determined by experiments. The fact that we are capable to identify slots i.e. a sequence of clock cycles that are dedicated to the processing of a single key bit is due to the fact that we could identify all other activities of the chips as well as thorough analysis of the recorded traces.

TABLE III. IMPLEMENTATIONS DETAILS OF THE ATTACKED CHIPS IN COMPARISON TO IHP *kP* ACCELERATOR

| *Parameter* | *IHP design, Spartan7 FPGA* | *Infineon Optiga Trust B SLE95250* | *NXP A1006* |
|---|---|---|---|
| Elliptic curve | NIST B-233 | given only: 131 bit key length [4] | NIST B-163 |
| Clock frequency | 100 MHz | 8 MHz | instable clock |
| *kP* execution time, ms (number of clock cycles) | 0.13 ms (<14 000) | 23 ms (up to 200 000*) | 37 ms (~17 053**) |
| Divisible into slots | yes | yes | yes |
| Implemented algorithm | Montgomery with LD | no information found | Montgomery with LD [32] |
| details about the algorithm | left-to-right | no details | no details |
| other implementation details | yes: white box | no details | no details |
| Attack success | 100% (the correctness of the best key candidate 98%, brute force for several bits) | less than 100%, exact evaluation without knowing the processed key is impossible | less than 100%, exact evaluation without knowing the processed key is impossible |

*Values calculated from frequency and *kP* execution time
**manually counted

When assessing the success rate of our attacks against the two commercial chips, the problem is that we cannot verify the actual number of key bits revealed correctly as we do not know the processed keys. So finally we can say we learned a lot about the behavior of the chips by our experiments, and that due to similar algorithms used in the implementation, there is a reasonable chance that our attack could reveal a significant number of key bits, if we would get the missing information. The investigated designs can be considered as still secure due to a kind of information hiding.

V. CONCLUSION

In this paper we reported our experiences of acting like a malicious attacker. In order to familiarize the reader with the attack procedure and the information needed to run a successful attack we provided details of our own implementation as well as of our own attack. We used our experience with attacks against different platforms to highlight the challenges when applying an attack to a more or less fully unknown device. While in case of our own implementation we did a kind of white box cryptography attacking the two commercial authentication chips, i.e. Infineon's Optiga Trust B SLE95250 and NXP's A1006 was real black box cryptography. The fact that only very limited information is available for both chips caused significant effort to learn at least the basics of the implementations. We even needed to extract relatively harmless information such as clock frequency and execution time by experiments. For the Optiga Trust B SLE95250 there is no information about the implementation available except the key length which seems to be the minimum to convince customers that the design provides a certain level of security. NXP provides more information at least the elliptic curve used is specified, and the mentioning of Lopez Dahab coordinates in [32] allows to deduce that NXP uses the Montgomery algorithm in its implementation. Please note that we did not find all this information in a one-stop-shop but needed to search the Internet to gather information or needed to run experiments, which is a significant effort. Despite all these challenges we were capable to identify all phases of the authentication i.e. initialization, communication, processing of the *kP* operation. Based on that knowledge we could analyse the *kP* operation and could successfully identify the processing of the key bits. As we do not know the secret key we could not determine how many of the key bits were determined correctly.

In the past we already attacked designs not implemented by us. But in those cases we received VHDL code so that our starting point was pretty close to the one when analyzing our own designs. Our experiments with the designs attacked here show that the effort for a successful attack without any knowledge about the implementation is about an order of magnitude higher. But in order to do a proper assessment of the security level of a certain implementation the procedures for certification for example according to CCS should be applied, as in that case all information needed to is provided by the manufacturer. We are stressing this point due to the essential differences between us and malicious attackers. The latter most probably would not hesitate to sing NDAs to retrieve confidential information which they then exploit for their attack, while we only used information that was publicly available. But when started working with the A1007 we learned that for the new chip NXP decided to provide even less data than for its predecessor, which in our opinion kind of indicate how sensitive NXP considers such information about its products and their security.

APPENDIX

X-Ray photos of a part of the OPTIGA Trust B Evaluation Kit board and SLE 95250 chip(IC U308).

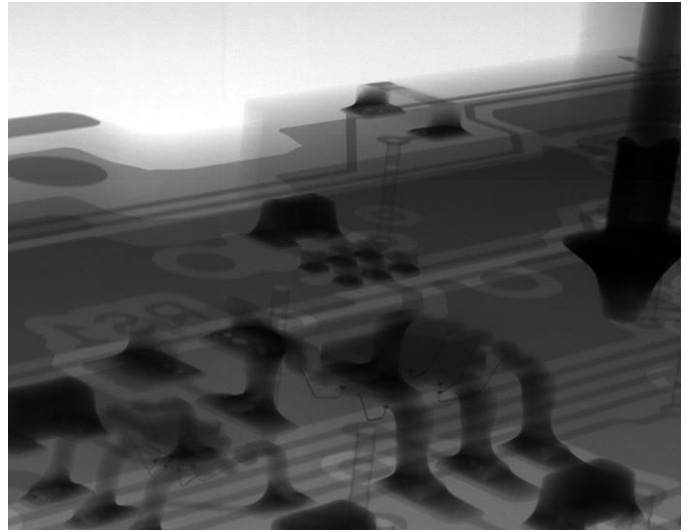

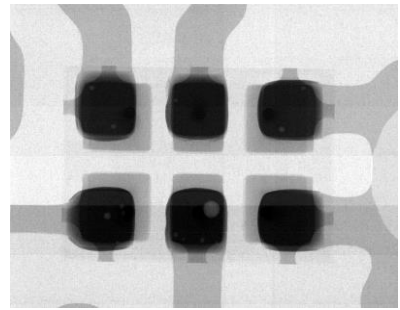

REFERENCES

[1] Neal Koblitz: "A Course in Number Theory and Cryptography". Second Edition. Springer-Verlag New York Inc., 1994
[2] Federal Information Processing Standard (FIPS) 186-4, Digital Signature Standard; Request for Comments on the NIST-Recommended Elliptic Curves: 2015. DOI http://dx.doi.org/10.6028/NIST.FIPS.186-4
[3] NXP A1006: Secure Authenticator IC https://www.nxp.com/products/identification-and-security/authentication/secure-authenticator-ic-embedded-security-platform:A1006
[4] Infineon OPTIGA™ Trust B SLE95250 Product Brief https://www.infineon.com/dgdl/Infineon-OPTIGA_Trust_B_SLE95250-PB-v01_00-EN.pdf?fileId=5546d4625b04ae11015b0f3f7f1c332e (last accessed 25.02.2019)
[5] I. Kabin, Z. Dyka, D. Klann, and P. Langendoerfer, "On the Complexity of Attacking Commercial Authentication Products," in 2019 10th IFIP International Conference on New Technologies, Mobility and Security (NTMS), Jun. 2019, pp. 1–6, doi: 10.1109/NTMS.2019.8763782.
[6] J. Lopez and R. Dahab, "Fast multiplication on elliptic curves over $GF(2^m)$ without precomputation," Proc. of CHES 1999, LNCS Vol. 1717, Springer, 1999.

[7] D. Hankerson, J. Lopez, and A. Menezes, "Software implementation of elliptic curve cryptography over binary fields," in Proc. of CHES 2000, LNCS Vol. 1965, Springer, 2000.

[8] Z. Dyka, P. Langendoerfer: *Area efficient hardware implementation of elliptic curve cryptography by iteratively applying Karatsubas method.* Proc. of the DATE 2005, Vol.3, pp: 70-75

[9] I. Kabin, Z. Dyka, D. Klann, and P. Langendoerfer, "Methods increasing inherent resistance of ECC designs against horizontal attacks," Integration, vol. 73, pp. 50–67, Jul. 2020, doi: 10.1016/j.vlsi.2020.03.001.

[10] Z. Dyka, E. A. Bock, I. Kabin, and P. Langendoerfer, "Inherent Resistance of Efficient ECC Designs against SCA Attacks," in 2016 8th IFIP International Conference on New Technologies, Mobility and Security (NTMS), 2016, pp. 1–5

[11] K. Itoh, T. Izu, and M. Takenaka, "Address-Bit Differential Power Analysis of Cryptographic Schemes OK-ECDH and OK-ECDSA," in Cryptographic Hardware and Embedded Systems - CHES 2002, Aug. 2002, pp. 129–143, doi: 10.1007/3-540-36400-5_11.

[12] I. Kabin, Z. Dyka, D. Kreiser, and P. Langendoerfer, "Methods for Increasing the Resistance of Cryptographic Designs Against Horizontal DPA Attacks," in Information and Communications Security, Dec. 2017, pp. 225–235, doi: 10.1007/978-3-319-89500-0_20.

[13] E. A. Bock and Z. Dyka, "Vulnerability assessment of an IHP ECC implementation," Accessed: Jul. 13, 2020. [Online]. Available: https://opus4.kobv.de/opus4-btu/frontdoor/index/index/docId/3490.

[14] Kabin, I., Kreiser, D., Dyka, Z., Langendoerfer, P.: FPGA Implementation of ECC: Low-Cost Countermeasure against Horizontal Bus and Address-Bit SCA. In: 2018 International Conference on ReConFigurable Computing and FPGAs (ReConFig). pp. 1–7 (2018). https://doi.org/10.1109/RECONFIG.2018.8641732.

[15] I. Kabin, Z. Dyka, D. Klann, and P. Langendoerfer, "Horizontal Attacks Against ECC: From Simulations to ASIC," in Computer Security, Cham, 2020, pp. 64–76, doi: 10.1007/978-3-030-42051-2_5.

[16] IHP – SiGe: C BiCMOS technologies. https://www.ihp-microelectronics.com/de/services/mpw-prototyping/sigec-bicmos-technologies.html

[17] Synopsys: DC Compiler, PrimeTime http://www.synopsys.com/Tools/

[18] B. L. Welch, "The generalisation of student's problems when several different population variances are involved," Biometrika, vol. 34, no. 1–2, pp. 28–35, 1947, doi: 10.1093/biomet/34.1-2.28.

[19] LANGER EMV-Technik GmbH. MFA02 micro probe set. https://www.langer-emv.com/de/product/mfa-aktiv-1-mhz-6-ghz/32/mfa-02-set-mikrosonden-1-mhz-bis-1-ghz/618/mfa-r-0-2-75-nahfeldmikrosonde-1-mhz-bis-1-ghz/854

[20] LANGER EMV-Technik GmbH. ICS 105 set - IC Scanner 4-Axis Positioning System. https://www.langer-emv.de/en/product/langer-scanner/41/ics-105-set-ic-scanner-4-axis-positioning-system/144

[21] Cmod S7: Breadboardable Spartan-7 FPGA Module https://store.digilentinc.com/cmod-s7-breadboardable-spartan-7-fpga-module/

[22] Digilent CmodS7 Schematic https://s3-us-west-2.amazonaws.com/digilent/resources/programmable-logic/cmod-s7/Cmod+S7_sch-public.pdf

[23] The Elliptic Curve Digital Signature Algorithm (ECDSA), http://cs.ucsb.edu/~koc/ccs130h/notes/ecdsa-cert.pdf

[24] J.-S. Coron, "Resistance Against Differential Power Analysis For Elliptic Curve Cryptosystems," in *Cryptographic Hardware and Embedded Systems*, Aug. 1999, pp. 292–302, doi: 10.1007/3-540-48059-5_25.

[25] Common Criteria for Information Technology Security Evaluation, Part 3: Security assurance components Version 3.1, Revision 5, April 2017. https://www.commoncriteriaportal.org/files/ccfiles/CCPART3V3.1R5.pdf

[26] Federal Office for Information Security, "Minimum Requirements for Evaluating Side-Channel Attack Resistance of Elliptic Curve Implementations", version 2.0, 2016, https://www.bsi.bund.de/SharedDocs/Downloads/DE/BSI/Zertifizierung/Interpretationen/AIS_46_ECCGuide_e_pdf.pdf;jsessionid=2FA5896575C424A7D03FC3331CA95FB8.2_cid351?__blob=publicationFile&v=3

[27] Lex Schoonen, "Overview of security evaluation", Brightsight presentation, available online (last viewed in April 2020): https://www.cosic.esat.kuleuven.be/ecrypt/courses/albena11/slides/lex_schoonen_security_evaluation.pdf

[28] Itoh, K., Izu, T., Takenaka, M.: A Practical Countermeasure against Address-Bit Differential Power Analysis. In: Cryptographic Hardware and Embedded Systems - CHES 2003. pp. 382–396. Springer, Berlin, Heidelberg (2003). https://doi.org/10.1007/978-3-540-45238-6_30.

[29] Izumi, M., Ikegami, J., Sakiyama, K., Ohta, K.: Improved countermeasure against Address-bit DPA for ECC scalar multiplication. In: 2010 Design, Automation Test in Europe Conference Exhibition (DATE 2010). pp. 981–984 (2010). https://doi.org/10.1109/DATE.2010.5456907.

[30] L. Batina, J. Hogenboom, N. Mentens, J. Moelans, and J. Vliegen, "Side-channel evaluation of FPGA implementations of binary Edwards curves," in 2010 17th IEEE International Conference on Electronics, Circuits and Systems, Dec. 2010, pp. 1248–1251, doi: 10.1109/ICECS.2010.5724745.

[31] Kabin, I., Dyka, Z., Klann, D., Langendoerfer, P.: Horizontal DPA Attacks against ECC: Impact of Implemented Field Multiplication Formula. In: 2019 14th International Conference on Design Technology of Integrated Systems In Nanoscale Era (DTIS). pp. 1–6 (2019). https://doi.org/10.1109/DTIS.2019.8735011.

[32] NXP Application note AN11875. Rev. 0.2 — 10 August 2016: A1006 Host Reference Implementation for LPC1115

[33] SOT1375-4. WLCSP4, wafer level chip-scale package. https://www.nxp.com/docs/en/package-information/SOT1375-4.pdf

[34] A1006 Secure Authenticator IC. Rev. 1 — 25 July 2018. Product short data sheet. https://www.nxp.com/docs/en/data-sheet/A1006-SDS.pdf

[35] ISO/IEC: 29167-12 Crypto suite ECC-DH security services for air interface communications. https://www.iso.org/standard/60442.html

[36] Yong Zhao, "Secure Tamper Resistant Authentication for Anti-Counterfeit Applications," September 2018, AMF-TRD-T3152. https://www.be1scrm.com/community/servlet/JiveServlet/downloadBody/341530-102-1-285990/APF-TRD-T3152_China.pdf

[37] J. Salvador, "Secure Tamper Resistant Authentication for Anti-Counterfeit Applications", June 2018, AMF-TRD-T3152. https://community.nxp.com/servlet/JiveServlet/downloadBody/340851-102-1-284788/AMF-TRD-T3152.pdf

[38] I. Galloway, "BMS for Drones and Small Systems, Incorporating Automotive Components", October 2019, Session #AMF-AUT-T3828. https://community.nxp.com/servlet/JiveServlet/downloadBody/344719-102-1-294341/AMF-AUT-T3828-RDDRONE-BMS772_rev.pdf